\documentclass[%
 reprint,
 amsmath,amssymb,
 aps,
]{revtex4-2}

\usepackage{graphicx}
\usepackage{dcolumn}
\usepackage{bm}
\usepackage{xcolor}

\newcommand\apjl{ApJ}                
\newcommand\mnras{MNRAS}             
\newcommand\aap{A\&A}                
\newcommand\apjs{ApJS}               
\newcommand\aj{AJ}                   
\newcommand\pasj{PASJ}               
\newcommand\jcap{J.~Cosmology Astropart. Phys.} 
\newcommand\ssr{Space Sci. Rev.}     
\newcommand\physrep{Reports on Progress in Physics}     
\newcommand\araa{Annual Rev. of Astronomy and Astrophysics}

\begin{document}

\preprint{APS/123-QED}

\title{Cosmic rays, neutrinos and GeV-TeV gamma rays from Starburst Galaxy NGC 4945}

\author{E. Aguilar-Ruiz$^{1}$}\email{E-mail: eaguilar@astro.unam.mx}
\author{N. Fraija$^{1}$}%
 \author{Jagdish C. Joshi $^{2,3}$}
  \author{A. Galvan-Gamez$^{1}$}
  \author{J. A. de Diego$^{1}$}
\affiliation{%
 $^{1}$Instituto de Astronom\'ia, Universidad Nacional Aut\'onoma de M\'exico, Circuito Exterior, C.U., A. Postal 70-264, 04510 M\'exico D.F., M\'exico\\
$^{2}$School of Astronomy and Space Science, Nanjing University, Nanjing 210093, China\\
$^{3}$Key laboratory of Modern Astronomy and Astrophysics
(Nanjing University), Ministry of Education, Nanjing 210093, China
}%

\date{\today}

\begin{abstract}
The detection of high-energy astrophysical neutrinos and ultra-high-energy cosmic rays (UHECRs) provides a new way to explore sources of cosmic rays. One of the highest energy neutrino events detected by IceCube, tagged as IC35, is close to the UHECR anisotropy region detected by Pierre Auger Observatory.  The nearby starburst galaxy (SBG), NGC 4945, is close to this anisotropic region and inside the mean angular error of the IC35 event. Considering the hypernovae contribution located in the SB region of NGC 4945, which can accelerate protons up to $\sim 10^{17} \, {\rm eV}$ and inject them into the interstellar medium, we investigate the origin of this event around this starburst galaxy.   We show that the interaction of these protons with the SB region's gas density could explain Fermi-LAT gamma-ray and radio observations if the magnetic field's strength in the SB region is the order of $\sim \rm mG$. Our estimated PeV neutrino events, in ten years, for this source is approximately 0.01 ($4\times10^{-4}$) if a proton spectral index of 2.4 (2.7) is considered, which would demonstrate that IC35 is not produced in the central region of this SBG.  Additionally, we consider the superwind region of NGC 4945 and show that protons can hardly be accelerated in it up to UHEs.
%
\end{abstract}

\maketitle

\section{\label{sec:level1}INTRODUCTION}
The detection of high-energy (HE) astrophysical neutrinos provides us the unique opportunity to investigate their origin. The IceCube observatory reported the discovery of 28 TeV-PeV neutrino events in 2013  \cite{2013PhRvL.111b1103A,2013Sci...342E...1I}.  In their latest update, 54 more neutrino events have been added to the high-energy starting events (HESE) catalog \cite{2017arXiv171001191I}. The Pierre Auger Observatory detected ultra-high-energy cosmic rays (UHECRs) above 57 EeV in their 3.7 years of observations \citep{2007Sci...318..938P, 2008APh....29..188P}. These observations exhibited an anisotropic distribution of UHECRs and their possible correlation with nearby extragalactic objects. Additionally, a large-scale anisotropic distribution was detected above 8 EeV, indicating a non-galactic origin \citep{2017Sci...357.1266P}.
The Pierre Auger Collaboration showed that if nearby starburst galaxies (SBGs) are included in the UHECR source population with Active Galactic Nuclei (AGNs), then patterns of arrival directions of UHECRs above 39 EeV could be explained with a $\rm 4 \sigma$ ($4.5\sigma$ for a recent update) confidence level against the isotropic hypothesis \citep{Aab_2018, thepierreaugercollaboration2019pierre}.
Furthermore, many works have pointed out that Centaurus A (a type of AGN)\citep{2012ApJ...746...72B,2009APh....31..138B, 2018MNRAS.481.4461F,2018MNRAS.479L..76M}, and/or M82 and NGC 253 (types of SBGs) \citep{1999PhRvD..60j3001A,2018PhRvD..97f3010A,Attallah_2018,2018A&A...616A..57R,2019arXiv190513243A} are potential UHECR candidate sources.
These astrophysical objects have also been proposed as candidates for HE neutrino production (for a review see \citep{2008PhR...458..173B, 2015RPPh...78l6901A}). Up to date,  TXS 0506 +056 is the only astrophysical object identified as a neutrino source  \cite{2018Sci...361.1378I, 2018Sci...361..147I}. However, a recent search for point-like neutrino sources showed a 2.9$\sigma$ fluctuation over the expected background near the location of the SB/Seyfert 2 galaxy NGC 1068 \cite{2020PhRvL.124e1103A}.  Although the significance is not enough to claim a real connection, several works have been developed to explore feasible explanations of the neutrino flux coming from this galaxy \cite{2021Galax...9...36I,2021arXiv210212409A,2021arXiv210204475K,2020ApJ...891L..33I}.
 
Inside the direction of the suggested Auger hotspot and the neutrino event IC35 with an energy of ${\rm 2004^{+236}_{-262} \; TeV}$, which is one of the most energetic neutrino events reported by IceCube Observatory in the HESE catalog, is located the galaxy NGC 4945. This galaxy is also detected in gamma rays by Fermi-LAT (Large Area Telescope) and reported in the first Fermi-LAT catalog \citep[1FGL;][]{2010ApJS..188..405A}. The gamma-ray emission was explained using the interaction of cosmic ray (CR) protons with the gas medium in the SB region \citep{2016ApJ...821...87E,2018MNRAS.474.4073W}.

NGC 4945 a nearly edge-on spiral galaxy classified as SB type is located at a distance of $\rm 3.82 \pm 0.31 \; Mpc$ \citep{1990MNRAS.245..665W,2007AJ....133..504K}. This galaxy is one of the brightest emitters in the infrared (IR) band, which originate from the heating of dusty medium by ultraviolet (UV) and optical photons from the central SB activity.   Moreover, optical line splitting provides evidence of an ionization region embedded within a conical cavity with a dimension of $\sim 1~ {\rm kpc} \times 6 \rm \; kpc$ along the minor and major axis of the galaxy \citep{1990ApJS...74..833H, 1989PASJ...41.1107N}. This conical-shaped region was also observed in the X-ray band, which is consistent with a super wind driven by SB activity \citep{10.1046/j.1365-8711.2002.05585.x}. Furthermore, NGC 4945 has an obscured AGN/Seyfert type 2 nucleus indicated by the variability on timescales of hours in the hard X-ray emission \citep{1993ApJ...409..155I}. This galaxy is considered one of the nearby brightest hard X-ray emitters at 100 keV only after NGC 4151 \citep{Done_1996}. The mass of its supermassive black hole is $\sim \rm 4 \times 10^6 \; M_\odot$, estimated by the detection of $\rm H_2O$ megamaser \citep{Greenhill_1997}.

\begin{figure}
		{
		\resizebox*{0.45\textwidth}{0.24\textheight}
		{\includegraphics{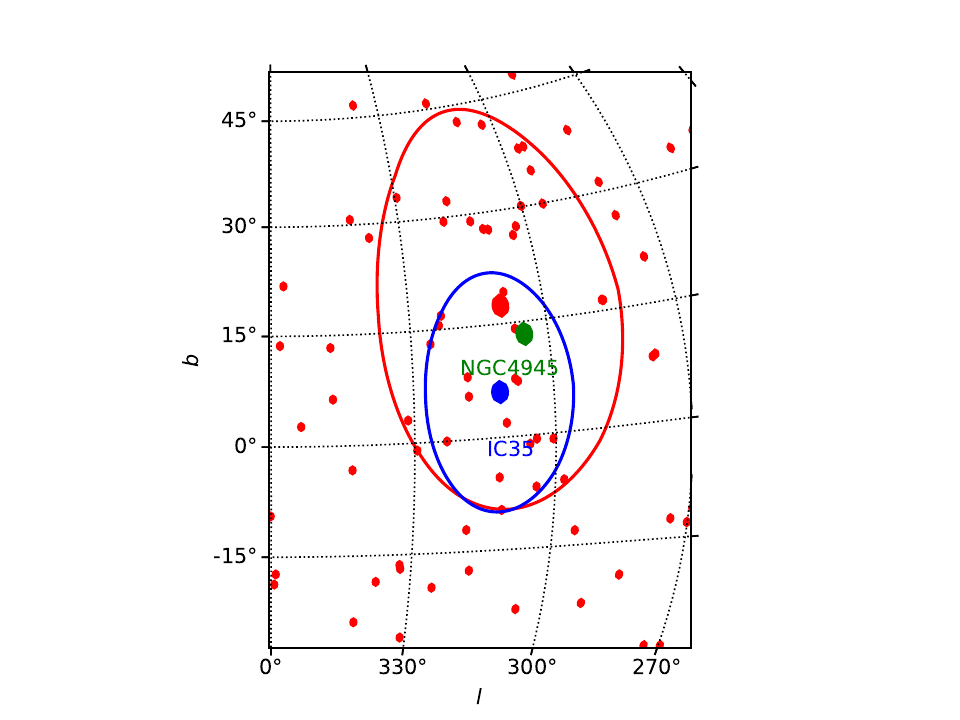}}
		}
		\caption{This figure shows the attractive region of the skymap, including the galaxy NGC 4945 (green point) and a region of $27^\circ$ centered at the biggest red-point (red circle) corresponding to Auger Hotspot \cite{thepierreaugercollaboration2019pierre}.  The neutrino IC35 corresponds to the blue point and its median angular error (blue circle). UHECRs are shown as red points.
		}\label{Fig_IC35_UHECRs}
\end{figure}

In this work, we investigate if there could be a connection among the SBG/Seyfert 2 NGC 4945, the IceCube neutrino event IC35, and UHECRs hotspot (Fig. \ref{Fig_IC35_UHECRs} shows their spatial correlation). We analyze if the nuclear SB region could produce PeV neutrinos via proton-proton ($pp$) collisions for protons accelerated up to $\sim 10^{17} \, {\rm eV}$. We assume that protons can be accelerated up to these energies by an energetic supernovae (SNe) explosion called hypernovae (HNe). This assumption differs from the work of  \cite{2018MNRAS.474.4073W} because they only consider SNe as proton acceleration sites, for which PeV neutrinos are not expected. Additionally, they did not consider the contribution of both primary and secondary pairs, which are important to explain the radio emission in the SB region. Although neutral pion decay products dominate the Fermi observations, the resulting cascade of secondary electrons could have significantly contribution. Furthermore, we estimate the maximum energy that CRs can reach in the superwind region and its possible contribution to the Auger hotspot.

The paper structure is as follows: in section 2, we introduce the sources of CRs in starburst galaxies, including two regions: the nuclear starburst and the superwind. Section 3 describes the theoretical model to calculate the gamma-ray and radio spectrum produced in starburst galaxies. Also, we include neutrino production. In section 4, we apply the model described in the previous section for the case of the starburst galaxy NGC 4945. Finally, in section 5, the conclusions are presented.
\section{Cosmic rays in SBG\lowercase{s}}
\subsection{Cosmic-ray acceleration requirements}
A charged particle moving along a uniform magnetic field has a helical motion with radius given by $r_L = E_{\rm CR}/\left( ZeB \right) \approx 1.08 \, Z \, E_{\rm 15, CR} B_{\mu G}^{-1} \rm \, pc$ where $Z$ is the atomic number, $E$ is the particle energy, $e$ is the electron charge, and $B$ is the strength of the magnetic field. The so-called  Hillas criterion provides the maximum energy that CRs can reach, expressed as $ r_L < R$ with $R$ the source's size. The maximum energy of CRs that a source can confine, independent of the acceleration mechanism, is given by
$ E^{\rm max}_{\rm CR} \approx 9.25 \times 10^{17} \; Z \, B_{\rm \mu G} \, R_{\rm kpc} \rm \; eV $
 \citep{doi:10.1146/annurev.aa.22.090184.002233}.
A useful parametrization of the characteristic acceleration timescale of CRs, independent of the Fermi first and second order acceleration mechanism is $ t_{\rm  acc} =  \mathcal{A} \, t_L$ with $\mathcal{A} \gtrsim 1$ \citep{2009JCAP...11..009L} and $t_{L}\simeq r_{L}/c$.  Therefore, the characteristic acceleration time can be written as
$ t_{\rm acc} \approx 350 \; Z^{-1} \, \mathcal{A} \, E_{\rm CR, 17} B_{\rm \mu G}^{-1} \rm \, yr,$
with the parameter $\mathcal{A}\propto \beta^{-2} D$, where $\beta$ is the velocity of the source and $D$ is the diffusion coefficient. This coefficient in the Bohm limit is $D_{\rm B} = 1/3 c \, r_L$, and in the case of Kolmogorov turbulence is $ D_{\rm K} \approx 1/3 c\,  r_L^{1/3}  l_c^{2/3}$ with $l_c$ the coherence length scale.

\subsection{Cosmic rays in SB region}
It is well accepted that CRs are accelerated in SN shocks via the Fermi mechanism up to energies of $\sim 10^{15}$ eV \citep{bell2004MNRAS550B}. Similarly, CRs can be accelerated up to higher energies in a special and most powerful SN type called HN. These HNe can provide CRs with energies as high as $\rm \sim 10^{17} \, eV$ \citep{PhysRevD.76.083009, He_2013, 2018SSRv..214...41B}. Therefore, CRs are injected by SNe and/or HNe into the SB region with a rate of $ Q(E) =  N \, E^{-\alpha}$, where $\alpha$ is the spectral index and the normalization constant $N$  in units of $\rm \left[ GeV^{-1}\, s^{-1}\right]$, is calculated using the total CR luminosity
\begin{equation}\label{eq:CR_luminosity}
L_{\rm cr} = \int_{E^{\rm min}}^{E^{\rm max}} E Q(E) dE = 
\left( f E_{\rm CR,  hn} + E_{\rm CR,  sn} \right) R_{\rm sn}
\; ,
\end{equation}
where $f= R_{\rm hn}/R_{\rm sn}$ is the ratio of HN to SN rates, $E_{\rm CR, hn} \left( E_{\rm CR, sn} \right)$ is the CR energy provided by HNe (SNe). We simplify the above equation defining the term in the parenthesis as the effective or average energy of CR supplied for both SNe or HNe
\begin{align}\label{eq:ECRav}
     \left< E_{\rm CR} \right> 
     &= \left( 1 + f \frac{\eta_{hn}E_{k, hn}}{ \eta_{\rm sn}E_{\rm k, sn}} \right)  \eta_{\rm sn}E_{\rm k, sn} \, ,
\end{align} 
where $\eta_{\rm hn} (\eta_{\rm sn})$ is the fraction of energy transferred from HN (SN) to CR acceleration, and $E_{\rm k,hn} (E_{\rm k, sn})$ is the typical energy released in a HNe (SNe). 
The terminology CRs refers to both electrons and protons. The proton density, $N_{\rm p}$, is higher in SBGs compared to the electron density, $N_{\rm e}$, ($\frac{N_{\rm p}}{N_{\rm e}} \sim 50$) \citep{Torres_2004}.  The steady-state of CR spectrum can be approximated as \citep{2014ApJ...780..137Y, 2013MNRAS.430.3171L}
\begin{equation}\label{eq:CR_steady}
N(E) \approx Q(E) \tau(E),
\end{equation}
where $\tau$ is the lifetime of CRs in the SB region.  CR distribution in the SB region is controlled by their diffusion in the random magnetic field and their outflow due to the galactic wind. The diffusion and advection time scales are defined as $t_{\rm diff} = 3h^2/(4D)$ and $t_{\rm adv} = h/V_{\rm w} $, respectively, where $V_{\rm w}$ is the galactic wind velocity, $h$ is the scale height of the SB region and $D$ is the diffusion coefficient.  The diffusion coefficient is parametrized as $D(E) = D_0 \, (E/{\rm 3 \, GeV})^{\delta}$, where $\delta$ lies in the range of $0.1\leq \delta\leq 1$ and depends on the spectrum of the magnetic turbulence of ISM. In SBGs, the Kolmogorov turbulence could be assumed, which agrees with recent studies of CR propagation in our galaxy with values of $\delta=0.3$ and  $D_0 \sim 6 \times 10^{28} \, \rm cm^2 \, s^{-1}$ \citep{2011ApJ...729..106T}.
However, the escape time will be a competition between diffusion and advection process and defined as $t_{\rm esc}^{-1} = t_{\rm diff}^{-1} + t_{\rm adv}^{-1}$.
The lifetime of CRs is compound by losses and escape, i.e,  $\tau = (t_{ \rm loss}^{-1} + t_{\rm esc}^{-1})^{-1}$. The Appendix lists the main loss processes for protons and electrons.
\subsection{Cosmic rays in Superwinds}

The SN-driven Superwinds have been investigated by \cite{1985Natur.317...44C, 2009ApJ...697.2030S}. The Superwind is expected when SNe and stellar winds collide each other. This collision forms a shock that thermalizes the central region and creates a cavity with heated gas at a temperature $T\sim 10^7\, {\rm K}$. This hot gas can reach the pressure necessary to produce an unbound gas under the gravitational potential, leading to an outflow away from the SB region. The escaping gas creates a region with an X-ray emission surrounded by a warm material detected in the optical band.
 
In Superwind theory, there are two main quantities, (i) the rate of energy transfer from SNe and stellar winds to the superwind, defined by $\dot{E} = \epsilon \dot{E}_*$  and (ii) the rate at which the hot matter is injected into the Superwind, i.e., $\dot{M} = \beta \dot{M}_*$. Using these quantities, we determine the temperature of hot gas
$
T_c = (\gamma_a-1) \mu m_p  \dot{E} / (\gamma_a\dot{M} k_B ) ,
$
where $k_B$ is the Boltzmann constant, $\gamma_a$ is the adiabatic index, and $\mu$ is the mean molecular weight. The number density of the hot gas in the central cavity is given by
$
n_c = 0.592 \sqrt{\dot{M}^3} / ( \sqrt{\dot{E}} \mu m_p (R + 2h)R)\, ,
$
where $R$ and $h$ are the radius and half-scale height of the nuclear star formation region, respectively. Therefore, the thermal pressure in the central cavity is $P_c = n_c k_B T_c$.
Once the wind escapes from the nuclear star forming region, the superwind undergoes re-acceleration, reaching a terminal velocity of
$
V_\infty = \sqrt{2 \dot{E} / \dot{M}}.
$
The material is dragged through the disk and the halo from the star formation region in a strong shock.  This strong shock accelerates particles via the Fermi mechanism producing a power-law distribution with a spectral index of $\alpha \sim 2$. The total CR luminosity provided by the superwind is $L_{\rm cr, sw} = \xi \dot{M} V_{\rm sw}^2$, where $\xi$ is the fraction of the Superwind energy used to accelerate particles and $V_{\rm sw}$ is the superwind velocity. 
The maximum energy reached in the superwind region can be estimated comparing $t_{\rm acc} \lesssim t_{\rm dyn}$ where $t_{\rm dyn} = R_{sw} / V_\infty$ is the dynamical timescale.
%
%
%
\subsection{UHECRs}
The recent anisotropic region reported by Pierre Auger Collaboration hints at a possible association with the starburst galaxy NGC 4945. This galaxy is 6° away from the hotspot's center, being the astrophysical object with a larger contribution in the model applied based on starburst galaxies \citep{2019ICRC...36..206C}.  This anisotropic region was parameterized with a circle centered at R.A. = $202^\circ$ and $\delta=-45^\circ$. This region has an excess of 62 events above the expected for the isotropic case. To estimate the expected number of UHECRs that could be observed from NGC 4945, $N_{\rm UHECR}$, as function of the bolometric luminosity of UHECRs, $L_{\rm cr}$, we use the relation given by \citep{2018MNRAS.481.4461F}
\begin{equation}\label{eq_L_UHECR}
    N_{\rm UHECR} = \frac{1}{4\pi d_L^2} \frac{(\alpha-2)}{(\alpha-1)} \frac{\Xi \omega(\delta_s)}{\Omega_{60} \, {\rm GeV}} \,  E_{\rm cr,min}^{\alpha-2} \, E_{\rm th}^{-\alpha+1} \, L_{\rm cr}
\end{equation}
where $\Xi \omega(\delta) / \Omega_{60} \simeq ( 101 400 \times  0.64 / \pi )\, \rm km^2 \, yr$ is the exposure of the Pierre Auger Observatory with almost 15 years of data, $E_{\rm cr, min}$ is the minimum energy of the CR spectrum and $E_{\rm th}> \rm 38 \, EeV$ is the threshold energy.
%
%
%
%
\section{Gamma rays and neutrinos from SBG\lowercase{S}}
\subsection{Gamma-ray production}

\subsubsection{Neutral pion decay products}
During their propagation, protons interact with the gas density of the medium, and the principal energy loss is via inelastic hadronuclear $(pp)$ collision \cite[e.g.,][]{2012ApJ...753...40F,2016ApJ...830...81F,2019JCAP...08..023F, }.  The collision timescale is given by $t_{\rm pp} \simeq(\kappa	 \,  c \, \sigma_{\rm pp} \, n_{\rm g})^{-1}$, where $\kappa \simeq 0.5$ is the inelasticity, $n_{\rm g}$ is the average gas number density of the medium and $\sigma_{\rm pp}$ is the inelastic cross section \citep{PhysRevD.74.034018,2014ApJ...783...44F},
\begin{equation}
    \sigma_{\rm pp}(E_{\rm p}) = (34.3 + 1.88L + 0.25L^2) \times \left[ 1 - \left( \frac{E_{\rm th}}{E_{\rm p}}\right)\right]^2 \; \rm mb,
\end{equation}
where $E_{\rm th}= 1.22 \, \rm GeV$ and $L=\ln \left(E_p /{\rm TeV}\right)$. 

The $pp$ collision produces neutral pions, and their production rate using $\delta$-approximation is given by \citep{PhysRevD.74.034018},
\begin{equation}
    q_\pi (E_\pi) = c \Tilde{n} \frac{n_{g}}{K_\pi} \sigma_{\rm pp} \left(m_\pi + \frac{E_\pi}{K_\pi} \right) \, N_p \left(m_\pi + \frac{E_\pi}{K_\pi} \right),
\end{equation}
where $\Tilde{n} \approx 1$, $K_\pi \approx 0.17$, $E_\pi$ is the pion energy and $m_\pi$ is the pion mass.
Neutral pions decay into two gamma rays \citep[$\pi^0 \rightarrow \gamma \gamma$;][]{2014MNRAS.437.2187F, 2014MNRAS.441.1209F, 2015MNRAS.450.2784F}. The gamma-ray spectrum is \citep{1968ApJ...151..881S}
\begin{equation}
    Q_\gamma^{pp}(E_\gamma) = 2 \int_{E_{min}}^\infty \frac{q_\pi (E_\pi)}{\sqrt{E_\pi^2 - m_\pi ^2 c^4}} dE_\pi,
\end{equation}
where $E_{\rm min} = E_\gamma + \frac{m_\pi^2 c^4}{4E_\gamma}$.
\subsubsection{Bremmstrahlung}
Another significant gamma-ray production is via Bremsstrahlung, for which the spectrum produced is given by \cite{Stecker1971}
\begin{equation}
    Q_{\gamma}^{\rm brem} (E_\gamma ) = n_g \, c \,  \sigma_{\rm brem} E_\gamma^{-1} \int_{E_\gamma/(m_e c^2)}^\infty N_e (\gamma_e) d\gamma_e \, \,,
\end{equation}
where $\sigma_{\rm brem} = 3.38 \times 10^{-26} \rm cm^{-2}$.
\subsubsection{Inverse Compton Scattering}
Ultra-relativistic electrons will suffer Compton scattering with the radiation field; lower-energy photons are scattered up to very high energies (VHEs). The spectrum produced by photons scattered, including the Klein-Nishina regime, is \citep{RevModPhys.42.237}
\begin{equation}
Q_\gamma^c (E_\gamma^c) = \frac{3}{4}c  \sigma_T \int d\gamma_e \frac{N_e(\gamma_e)}{\gamma_e^2}  \int d\epsilon \frac{n_{ph}}{\epsilon} F_c(q,\Gamma_e) \, ,
\end{equation}
where $F_c$ is the Compton scattering kernel for an isotropic photon and electron distributions with $q=\frac{E_\gamma^c}{\Gamma_e \left( \gamma_e m_e c^2 - E_\gamma^c \right)}$ and $\Gamma_e = \frac { 4 \epsilon \gamma_e } { m_e c^2}$.

\subsection{Secondary electrons}
\subsubsection{Pionic $e^\pm$ production}
Electrons produced by $pp$ collision can be approximated as the pion production rate because the muon moves nearly the pion speed. Then their source functions can be equivalent $q_\mu(\gamma_\mu) \simeq q_\pi(\gamma_\pi)$ with a little bit difference on the value of $\Tilde{n} \approx 0.77 ; 0.62 ; 0.67$ for spectral index $\alpha = 2 ; 2.5 ; 3$ \cite{PhysRevD.74.034018}. Then, the electron production rate is given by
\begin{equation}
Q_e^{pp}(\gamma_e) = \int_1^{\gamma'_{e, \rm max}} d\gamma_e' \frac{P(\gamma_e')}{2\sqrt{{\gamma'_e}^2 - 1}} \int_{\gamma_\mu^{-}}^{\gamma_\mu^{+}} d\gamma_\mu \frac{q_\mu(\gamma_\mu)}{\sqrt{\gamma_\mu^2 - 1}},
\end{equation}
where $\gamma_\mu^{\pm} = \gamma_e \gamma_e' \pm \sqrt{\gamma_e^2-1} \sqrt{{\gamma'_e}^2 - 1}$, $\gamma'_{e,\rm max}=104$, and the electron distribution in the muon rest frame is given by \cite{Schlickeiser2002}
\begin{equation}
P(\gamma_e') = \frac{2{\gamma'_e}^2}{{\gamma'_{e, \rm max}}^3} \left( 3 - \frac{2{\gamma'_e}}{{\gamma'_{e, \rm max}}} \right)\,.
\end{equation}
\subsubsection{$\gamma\gamma \rightarrow e^\pm$ production} 
The radiation field of SBGs will attenuate VHE gamma rays, where the principal internal attenuation source is the IR radiation. The optical depth is calculated assuming a homogeneous and isotropic photon distribution as \citep{1992ApJ...390L..49S, 10.2307/j.ctt7rphp,2020MNRAS.497.5318F}
\begin{equation}\label{eq_tau_eCreation}
    \tau(E_\gamma) = L \int_{-1}^{+1} d\mu \frac{(1-\mu)}{2} \int_{\epsilon_{th}}^\infty d\epsilon \; \sigma_{\gamma \gamma}(\beta) \; n_{\rm ph}(\epsilon)\,,
\end{equation}
where $L$ is the mean distance traveled by gamma rays, $\sigma_{\gamma \gamma}(\beta)$ is the total cross-section, $\beta= \left[ 1 - \frac{2 m_e c^2}{E_\gamma \epsilon (1-\mu)}\right]^{1/2}$ is the velocity of the pair created in the center-of-mass system, $\epsilon_{th} = \frac{2 m_e^2 c^4}{E_\gamma (1 - \mu)}$ is the threshold energy for the production of electron-positron pairs and $\mu$ is the cosine of the angle of collision. Assuming a planar geometry, the absorption coefficient of gamma rays in the SB region would be $C_{\rm abs}^{\rm SBs} (E_\gamma) = \left( 1- \exp^{-\tau_{\gamma\gamma}(E_\gamma)} \right) /\tau_{\gamma\gamma}(E_\gamma)$. Once gamma rays escape from the source, their absorption due to the extragalactic medium becomes important so that we include this effect using $C_{\rm abs}^{\rm EBL} = \exp^{-\tau_{\gamma\gamma}(E_\gamma)}$.
The pair production rate in the nuclear SB region  is \citep{2011ApJ...728...11I}

\begin{equation}
Q_e^{\gamma\gamma}(\gamma_e) = 2 \frac{dE_\gamma}{d\gamma_e} Q_\gamma(E_\gamma) \left[ 1 - C_{\rm abs}^{\rm SBs}(E_\gamma) \right]\,,
\end{equation}
where $E_\gamma \approx 2 \gamma_e m_e c^2$.

Finally, the observed gamma-ray spectrum due to hadronic, cascade processes and Bremsstrahlung is calculated using $Q_\gamma^{\rm obs} = \left( Q_\gamma^{pp} +  Q_{\gamma}^{c} +Q_\gamma^{\rm brem}\right) C_{\rm abs}^{\rm SBs}C_{\rm abs}^{\rm EBL}$.

\subsection{Radio emission}
Synchrotron emission is the main responsible for radio observation, which has an emissivity given by
\begin{equation}
    J_{\rm syn} (\epsilon) = \frac{\sqrt{3} e^3 B}{ 2 \pi \hbar m_e c^2}  \int_{\gamma_{e,min}}^{\gamma_{e,max}} d\gamma_e \, N_e(\gamma_e) \, R(x)\,,
\end{equation}
where $ x= 2 m_e \epsilon c / ( 3 e \hbar B \gamma_e^ 2)$ and the function $R(x)$ is defined in \cite{2008ApJ...686..181F}.
\subsection{Neutrino production} SB galaxies have been proposed as neutrino sources. The main assumption is that SNe accelerate protons and inject them into ISM shooting pion-production via hadronuclear interactions \citep{2006JCAP...05..003L}. As discussed by \cite{2013PhRvD..88l1301M, 2015ApJ...806...24S, He_2013,Xiao_2016}, HNe inside SB galaxies can provide CRs in order to explain ${\rm \,PeV}$ neutrinos.
 
A simple way to relate the CRs and neutrino production is via the efficiency of the process given by $f_\pi = 1- \exp \left(- t_{esc} / t_{pp} \right)$. Therefore, the relation between protons and all neutrino flavor in the case of $pp$ interactions is given by \citep{10.1093/ptep/ptx021},
\begin{equation}
E_\nu^2 Q_\nu(E_\nu) \simeq 0.5 f_\pi E_p^2 Q_p(E_p)\,.
\end{equation}
The average fraction of energy transferred from protons to neutrinos is $E_\nu \approx 0.05 E_p$. The expected number of neutrinos between 30 TeV and 10 PeV observed in the IceCube detector can be estimated from neutrino flux as
\begin{equation}\label{eq_Nnu}
N_\nu \approx \frac { T_{\rm obs} }{4 \pi D_L^2}\int_{\rm 30 \, TeV}^{\rm 10 \, PeV} dE_\nu \, A_{\rm eff} \left( E_\nu\right) Q_\nu(E_\nu), 
\end{equation}
with $T_{\rm obs}$ the observation time, $A_{\rm eff}$ the effective area at specific energy and $D_L$ the luminosity distance.


\section{Application: NGC 4945}

\subsection{Starburst region}
In the inner region ($R \sim$ 250 pc) for NGC 4945, \cite{Lenc_2008} derived the limits for the SN rate as $\rm 0.1 < (R_{SNe}/yr^{-1}) < 14.4 $, and the star formation rate as $\rm 2.4 < SFR/ \left( M_\odot \, yr^{-1} \right) < 370$. Similarly, \cite{10.1093/mnras/stw1659} obtained a value of $\rm SFR \sim 4.35 \, M_\odot \, yr^{-1}$ using $\rm H42\alpha$ and 85.69 GHz free-free emission. The total IR emission could also trace both the dusty region or the star formation region. Given this fact, \cite{Strickland_2004} reported a value of $\rm SFR \approx 4.6 \, M_\odot \, yr^{-1}$ using the total IR luminosity that arises in a region of $\rm 12" \times 9" (215 \times 161 \, pc)$ \citep{1988ApJ...329..208B}.
We follow the relation between the SFR and the gas contained in starburst galaxies, $\Sigma_{\rm SFR} \propto \Sigma_{g}^{1.4}$ \citep{Kennicutt_Jr__1998}, where $ \Sigma_{g} = 2h \mu \, m_p \, n_{g} $ is the surface gas density asuming a disk geometry. Then,  it is possible to estimate the gas content in the star formation region

\begin{multline}\label{eq:KS_ngas}
    n_{g} \approx 5 \times 10^3 \; \left( {\rm SFR / 4.6\, M_\odot \, yr^{-1} } \right)^{0.7}\, \times
    \\
    \left( R / 110 \, {\rm pc} \right)^{-1.43}\, 
    \left( h/80 \, {\rm pc} \right)^{-1} \, \rm cm^{-3}\,. 
\end{multline}
This value is in agreement with the estimation, $n_{\rm H_2} \approx (3-10) \times 10^3 \, \rm cm^{-3}$, obtained by \cite{2001A&A...367..457C}. 

\subsubsection{Magnetic fields}\label{section:Magnetic_field}
The magnetic field, $B$, inside a starburst region plays an important role to control diffusion and emission processes (e.g., synchrotron and IC). The strength's value will be reflected in the shape and intensity of the observed spectrum. For ultra-relativistic electrons, lower values of $B$ suppress synchrotron emission and lead to IC scattering to be the dominant process. Otherwise, values of B larger than $B_{\rm cut}$ (for which $t_{\rm syn} \approx t_{\rm IC}$), synchrotron emission becomes dominant, and therefore IC scattering is suppressed. The value of $B_{\rm cut} \approx 450 \, \rm \mu G$ is found with the condition $U_B \approx U_{\rm ph}$, where $U_{\rm ph} \approx \pi R^2 L_{\rm ph} \approx 7.36 \; \times 10^{-9} \, \rm erg \, cm^{-3}$ is the IR energy density and $U_B=B^2/(8\pi)$ is the magnetic energy density.  We estimate $B$ using different methods: (i) the relation with $\Sigma_{SFR}$ following \cite{2014ApJ...793..131C} (see, references therein) we have different values
\begin{equation}\label{eq:B_SFR}
B \approx \,
    \begin{cases}
        214 \, \left( {n_{g} / 10^{3.7} \, \rm cm^{-3}} \right)^{0.4}\, \left( h/ 10^{1.9} \, \rm pc \right)^{0.4} \; \rm \mu G
        \\
        750 \, \left( {n_{g} / 10^{3.7} \, \rm cm^{-3}} \right)^{0.7}\, \left( h/ 10^{1.9} \, \rm pc \right)^{0.7}\; \rm \mu G
        \\
        4900 \, \left( {n_{g} / 10^{3.7} \, \rm cm^{-3}} \right)\, \left( h/ 10^{1.9} \, \rm pc \right) \; \rm \mu G.
    \end{cases}
\end{equation}
(ii) The equipartition, $B_{\rm eq}$, and minimum, $B_{\rm min}$, values were previously estimated by \cite{2013MNRAS.430.3171L} under different assumptions. They found the ranges of $B_{\rm eq}=(110-130) \, \mu \rm G $ and $B_{\rm min}=(89-130) \, \mu \rm G $. 
(iii) Radio emission is a feasible way to set the value of $B$ if SF activity is assumed as its origin. Here, we estimate $B$ from radio observations by assuming that all emission in this band comes from primary and secondary electrons confined in the SB region.
%

\subsubsection{Gamma-ray absorption}

\begin{figure}
		{
	    	\resizebox*{0.5\textwidth}{0.28\textheight}
				{\includegraphics{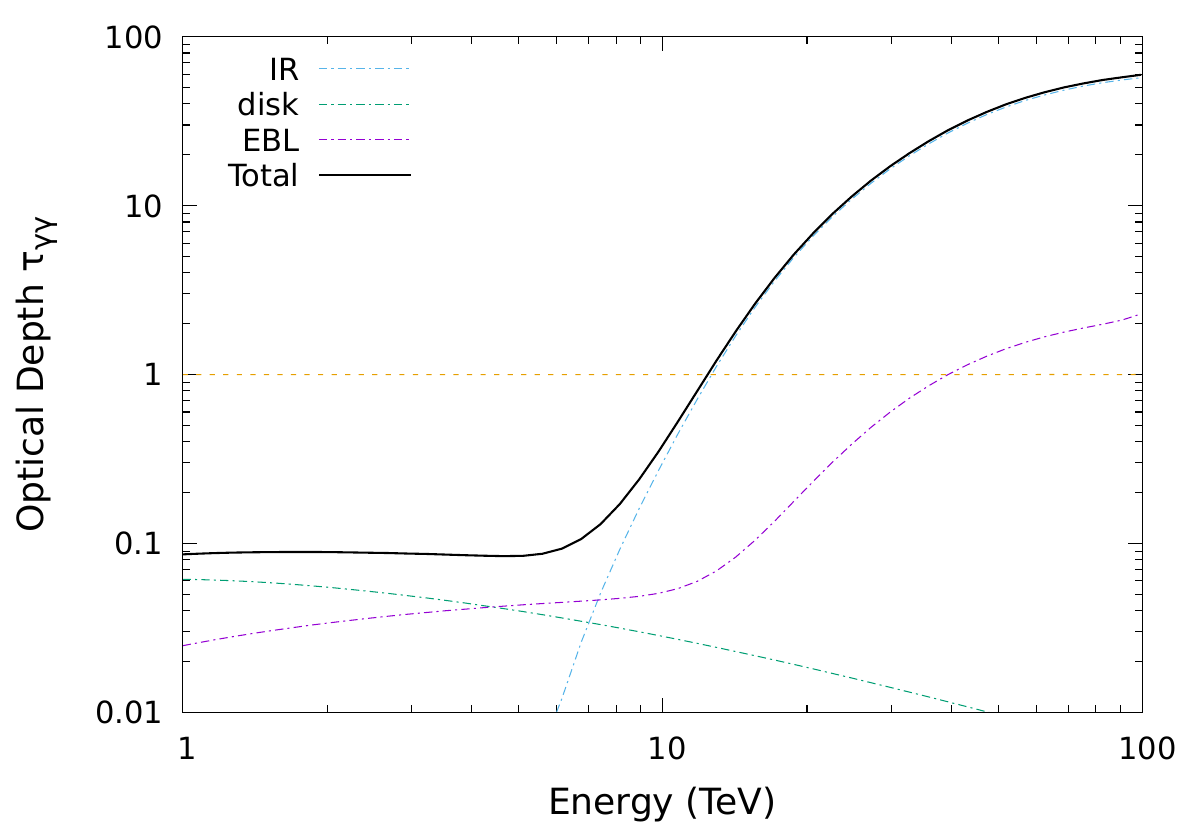}}
		}
		\caption{Total optical depth due to pair production in the SB region's radiation field: the IR radiation, disk radiation, and the external radiation field due to the extragalactic background.}
		\label{Fig_opdep}
\end{figure}

The pair production process strongly attenuates VHE gamma rays produced in the SB region due to the high density of IR photons. Taking the energy peak of IR spectrum provided by \cite{10.1093/mnras/stw1659}, the attenuation threshold for VHE gamma-rays in a head-on collision with IR photons is
$E_\gamma \approx 26 \left( \epsilon_{\rm IR} / \rm 0.01 \, eV \right)^{-1} \rm \; TeV.$
Additionally, \cite{Lenain_2010} discussed the existence of an accretion disk to explain the observed spectrum. This accretion disk radiation permeates the central region producing attenuation of gamma rays. Considering the energy corresponding to the peak flux of the model proposed in \cite{Lenain_2010}, the threshold energy of gamma rays for attenuation is
$E_\gamma \approx 0.87 \left( \epsilon_{\rm disk} / \rm 3 \, eV \right)^{-1} \rm \; TeV.
$ Finally, external attenuation due to the extragalactic background light (EBL) is considered using the parametrization introduced in \cite{2007arXiv0711.2804D}. 
The total optical depth is calculated using Eq. (\ref{eq_tau_eCreation}) and displayed in  Fig. \ref{Fig_opdep}. The result shows that the principal attenuation source above $\sim \rm 10 \, TeV$ is the IR radiation field. 
%
%
%
%
\subsubsection{Radio and gamma-ray spectrum}
We explain radio observation with synchrotron emission of electrons (primaries and secondaries) and gamma-rays with the hadronic model ($pp$ collisions). We consider secondary electrons, as pointed by  \citep{2013MNRAS.430.3171L}. They showed that these electrons are dominant over the primary ones by a factor $Q_{\rm e,sec}/ Q_{\rm e,prim} \approx 4.6$, for typical values of $\alpha = 2.2$, $N_p/N_e=50$ and $F_{cal}\approx 1$.

\begin{figure}
		{
        \resizebox*{0.5\textwidth}{0.28\textheight}
		{\includegraphics{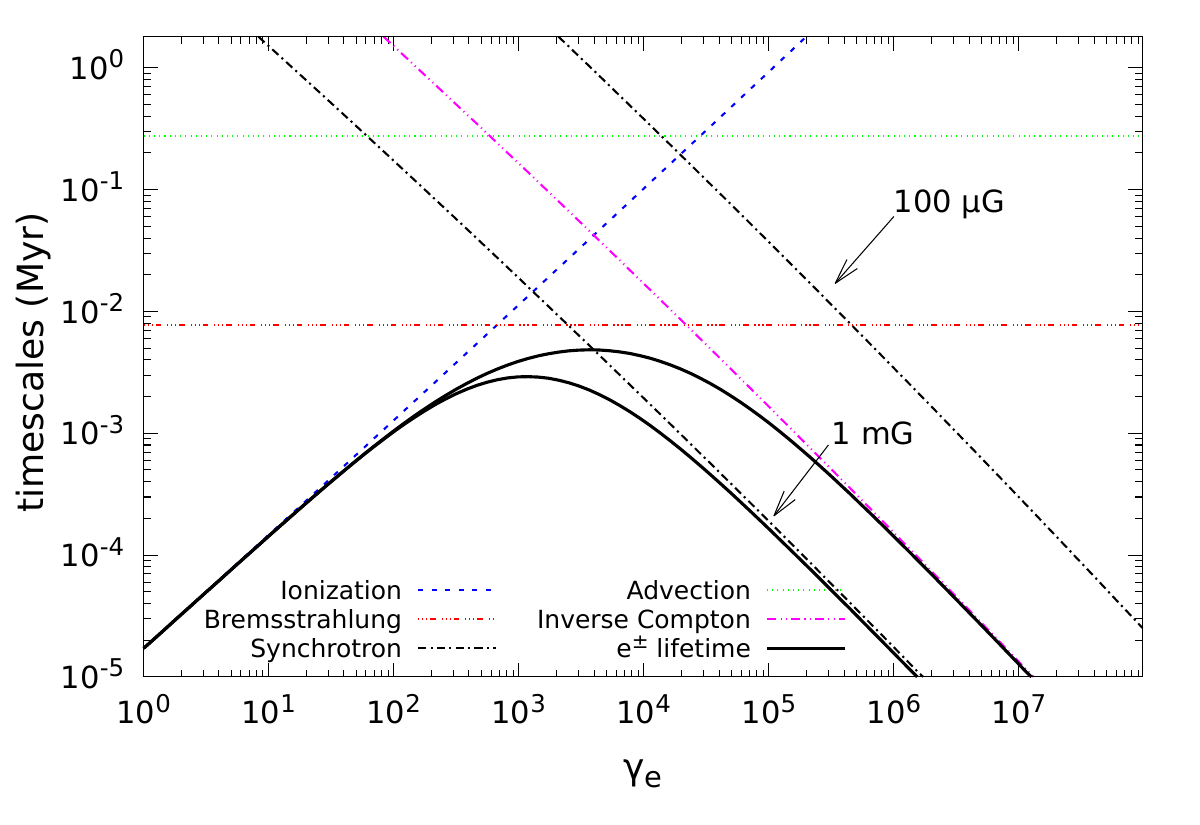}}
		}
		\caption{CR electron timescales: $e^\pm$ energy losses via ionization, Bremsstrahlung, inverse Compton and synchrotron process in the magnetic field of value $B=100 \, \rm \mu G$ and $B=1 \, \rm m G$, while escape is only due to advection (diffusion is neglected).}\label{fig_e_timescales}
\end{figure}

\begin{figure}
		{
			\resizebox*{0.5\textwidth}{0.28\textheight}
				{\includegraphics{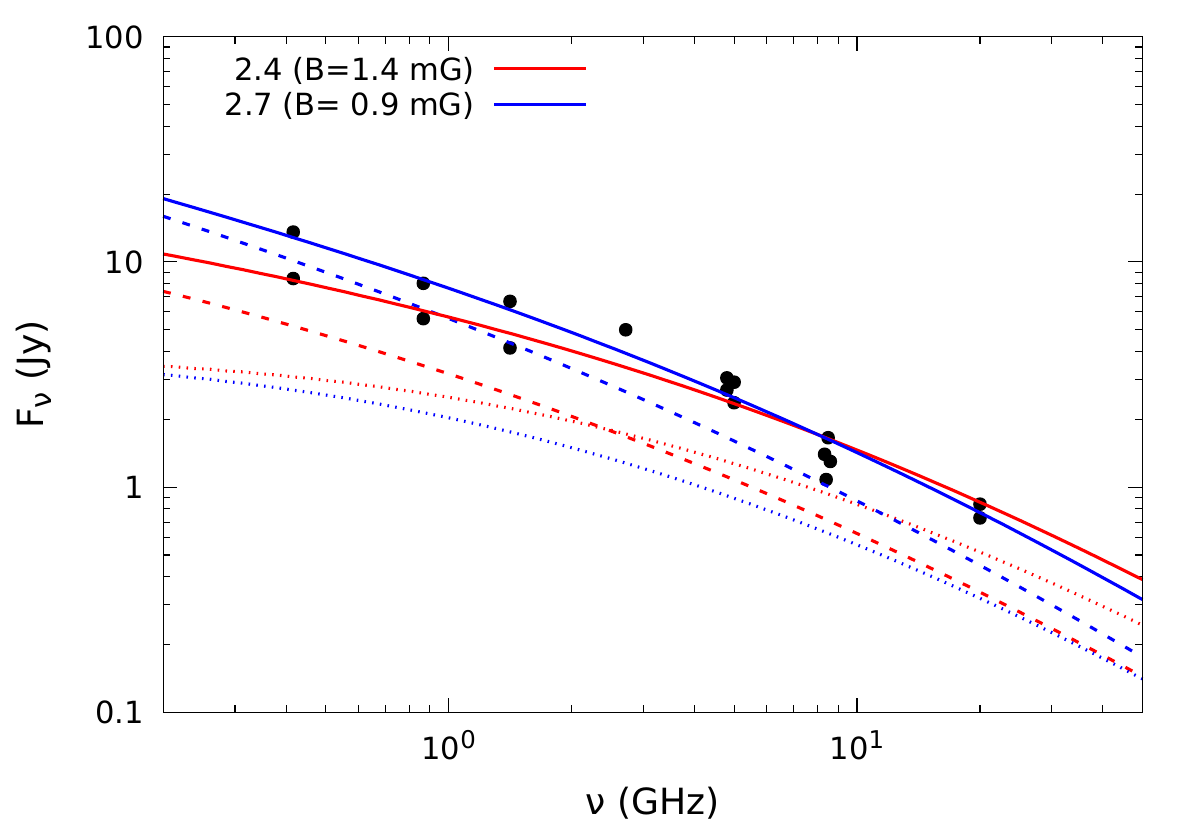}}
			}
		\caption{Spectral energy distribution for radio emission. The total flux in the central SB for spectral index 2.4 (red solid line) and 2.7 (blue solid line) is produced by primary (dotted-line) and secondary (dashed-line) electrons. We observed that if the spectral index is steeper, we demand a stronger magnetic field to explain the data. Data are taken from  \cite{10.1093/mnras/stw1659, Lenain_2010}  }\label{Fig:SED_syn}
\end{figure}

\begin{figure}
		{
		\resizebox*{0.5\textwidth}{0.28\textheight}
				{\includegraphics{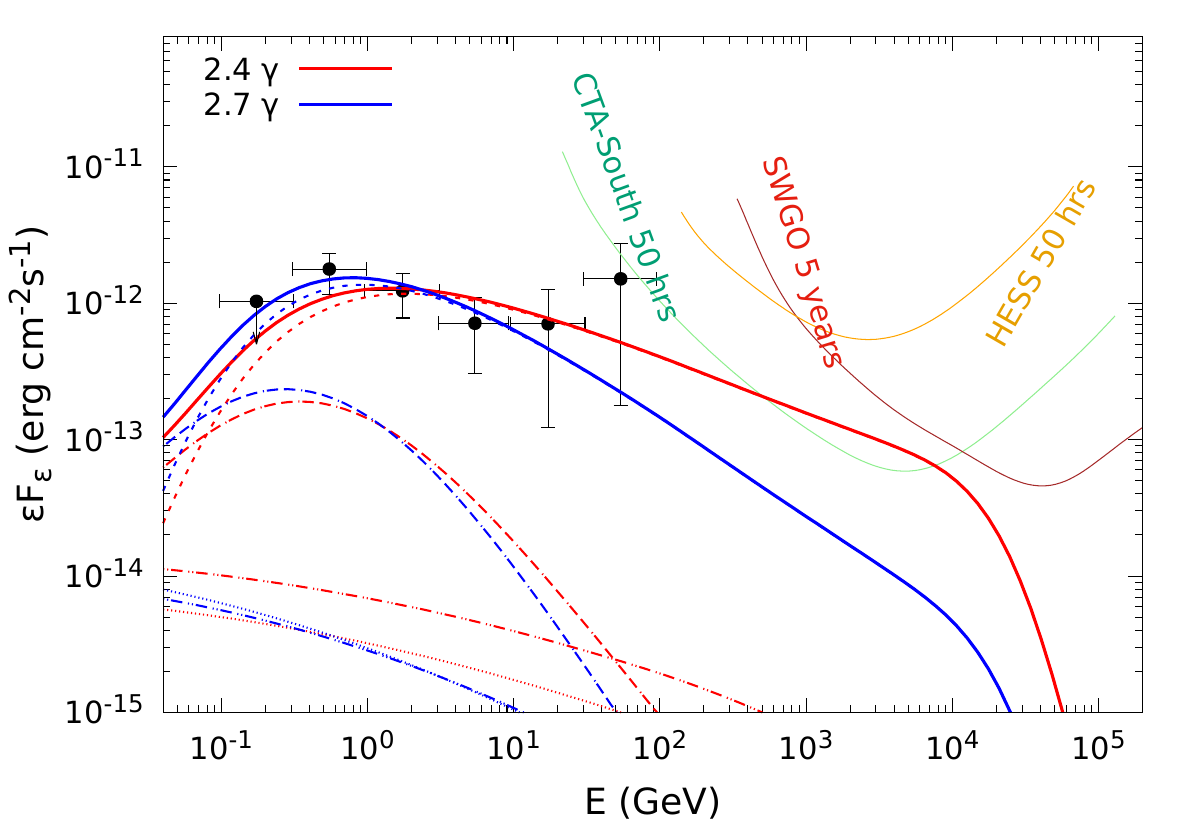}}
			}
		\caption{Spectral energy distribution for $\gamma$-rays. The total flux in the central SB region for spectral index 2.4 (red solid line) and 2.7 (blue solid line) is shown. The contribution from $pp$-model (dashed line), Bremsstrahlung (dotted-dashed-dashed line), external IC for secondary (dotted-dotted-dashed line), and primary (dotted line) are shown. The contribution from $\gamma\gamma$ absorption has no significant contribution and is not plotted.}\label{Fig_SED}
\end{figure}

We use the bolometric CR luminosity injected by SN (HN) and the spectral index as parameters (see Eq. \ref{eq:CR_luminosity}) to normalize the CR injection rate, $Q \propto E^{-\alpha}$.  Another parameter is the magnetic field, which influences mainly in the radio emission. In order to explain the radio emission with SF activity, $B\sim \rm mG$ is demanded. As discussed above, if $B \gtrsim 450 \, \rm \mu G$ electrons are cooled mainly by synchrotron while IC scattering is suppressed.
Additionally, we assume an equal spectral index for protons and primary electrons and a constant ratio between them $N_p/ N_e \sim 50$ \citep{Torres_2004}.  Finally, the steady-state spectrum, $N(E)$, was calculated using the lifetime inside the star formation region, as given by Eq. (\ref{eq:CR_steady}). The  computed lifetime of electrons and protons is plotted in Figs. (\ref{fig_p_timescales}) and (\ref{fig_e_timescales}), respectively.
In this work, we chose two extreme values $\alpha = 2.4$ and $2.7$, because softer and steeper values cannot fit the spectrum of gamma-rays appropriately (see Fig. \ref{Fig_SED}). We mention that only frequencies below $\lesssim 50 \, \rm GHz$ was considered, because at high energies free-free and dust emission are expected to be dominant. Therefore, the radio and gamma-rays spectra have good fits for $\alpha=2.4$, $L_{cr} = 1.3 \times 10^{42} \, \rm erg \, s^{-1}$ and $B = 1.4 \rm \, mG$ whereas if $\alpha=2.7$ is chosen, we demand  $L_{cr} = 1.82 \times 10^{42} \, \rm erg \, s^{-1}$ and $B = 0.9 \rm \, mG$. Results are plotted in Figures \ref{Fig:SED_syn} and \ref{Fig_SED}.   From our resulting radio spectrum, we observe that below frequencies of $\lesssim \rm GHz$ secondary electrons are the dominant population, but at high frequencies, primaries equal or even overwhelm the secondary contribution. In the case of $\gamma$-rays, our result shows that secondary electrons, although the dominant population over the primary, do not have essential contributions.  This does not happen for Bremmstrahlung because, at sub-GeV, energies cannot be neglected. Furthermore, depending on the choice of spectral index, future TeV gamma-ray observations are expected. As we observe in Fig. (\ref{Fig_SED}) only $\alpha\sim 2.4$ implies future detection by the Cerenkov Telescope Array (CTA)\citep{2019scta.book.....C}.

\subsubsection{Energy injection of Cosmic rays by HN and SN}

We consider the contribution of HN and SN to the injection of CRs. These CRs correspond to the source of the observed resulting spectrum. Until now, we do not emphasize what provides more CRs energy into the SB region; only the average CRs provided by both SN and HN can be estimated $\left< E_{\rm CR} \right> = L_{cr}/R_{sn}$ (see, Eq.\ref{eq:CR_luminosity}). Then, using our result from previous subsection we have, $\left< E_{\rm CR} \right> \approx 4.2 \times 10^{50} \rm \, erg$ and $\approx 5.9 \times 10^{50} \rm \, erg$ for $\alpha = 2.4$ and $2.7$, respectively. Hereafter, we use the value $\left< E_{\rm CR} \right> \approx 5 \times 10^{50} \rm \, erg$.

 Similar CRs luminosity from HNe and SNe must be expected because of HN rate represents a tiny fraction of the SN rate about of the 7\% \citep{Guetta_2007} but HNe are more energetic event with typical kinetic energy of $E_{ k,hn} = 10^{52} \, {\rm erg}$ whereas SNe have $E_{ k,sn} = 10^{51} \, {\rm erg}$ . In the case of NGC 4945 we have $R_{ sn} \gtrsim 0.1 \, \rm yr^{-1}$  (see Table \ref{tab:ObsParameters}) then will have $R_{ hn} \gtrsim 0.007 \, \rm yr^{-1}$. The  efficiency of CR acceleration of SNe, $\eta_{sn}$, is uncertain but some works based on the observation of our galaxy suggest that it lies in the range of $(0.1-0.3)$ (e.g., \cite{1990cup..book.....G, Caprioli_2012}). Note that with these values SNe cannot be the only accelerator of CRs because they demand kinetic energies  $E_{k, sn} = \left< E_{\rm CR} \right>/\eta_{sn} \approx (1.7-5)\times 10^{51} \, \rm erg$, which are slightly bigger than the typical SN kinetic energy. Therefore, the efficiency of CR acceleration by HNe could be estimated from Eq. (\ref{eq:ECRav}) as
\begin{multline}\label{eq:HN_CReff}
    \eta_{hn} = 0.28 \, \left( f/0.07 \right)  \left( \mathcal{C}/0.2\right) 
    \times \\
     \,\left( E_{k,sn} / 10^{51} \, {\rm erg} \right)
    \left(E_{k,hn}  / 10^{52} \, {\rm erg} \right)^{-1},    
\end{multline}
with
\begin{equation}\label{eq:Cvalue}
    \mathcal{C} = \left[  0.5 \left( \left< E_{\rm CR} \right>/ 10^{50.7} \, {\rm erg}\right)  \left( E_{k,sn} / 10^{51} \, {\rm erg} \right)^{-1} - \eta_{sn} \right] > 0\,.
\end{equation}
 Taking $\eta_{sn}=0.1-0.3$ and from Eq. (\ref{eq:Cvalue}), we have that the value of $\mathcal{C}$ lies in the range of $\approx 0.4 - 0.2 $.   If an HN has similar or less efficient particle acceleration than SN, therefore, only the higher value $\eta_{sn} \approx 0.3 (\mathcal{C}\approx0.2)$ leads to a feasible value of $\eta_{hn} \approx 0.28$. This suggests similar contributions of CRs from SNe and HNe to the SB region of NGC 4945, $L_{\rm cr,sn} = \eta_{sn} R_{ sn} E_{k,sn} = 8.87 \times 10^{41} \, \rm erg \, s^{-1}$ and $L_{\rm cr,hn} = \eta_{hn} R_{ hn} E_{k,hn} = 6.97 \times 10^{41} \, \rm erg \, s^{-1}$.

It is worth noting that we use the lower limit on $R_{\rm sn}$, which translates into an upper value of $\left< E_{\rm CR} \right> =L_{cr} / R_{sn} \lesssim 5 \times 10^{50} \, \rm erg$, but higher values of $R_{sn}$ can be considered, as estimated by \cite{Lenc_2008}.  However, examining Eq. (\ref{eq:Cvalue}) again, we notice that only $\left< E_{\rm CR} \right> > (1-3) \times 10^{50} \, \rm erg$ satisfies $C>0$ if typical values are chosen. This lower limit could be interpreted as a case of complete CR injection by SNe without any contribution by HNe. The last lower value set an upper limit on $R_{sn}< 1 \, \rm yr^{-1}$. Using the relationship given by \cite{1992ARA&A..30..575C}, we obtain ${\rm SFR} < 24.4 \, \rm M_{\odot} \, yr^{-1}$. Note that this result suggests that SN rate and SF rate cannot be as high as the values estimated by \cite{Lenc_2008}. Our result of the SB region is summarized in the Table \ref{tab:DerivedParameters_SB}.

\subsubsection{Expected HE-neutrino and UHECR Events}

\begin{figure}
		{
    	\resizebox*{0.5\textwidth}{0.28\textheight}
				{\includegraphics{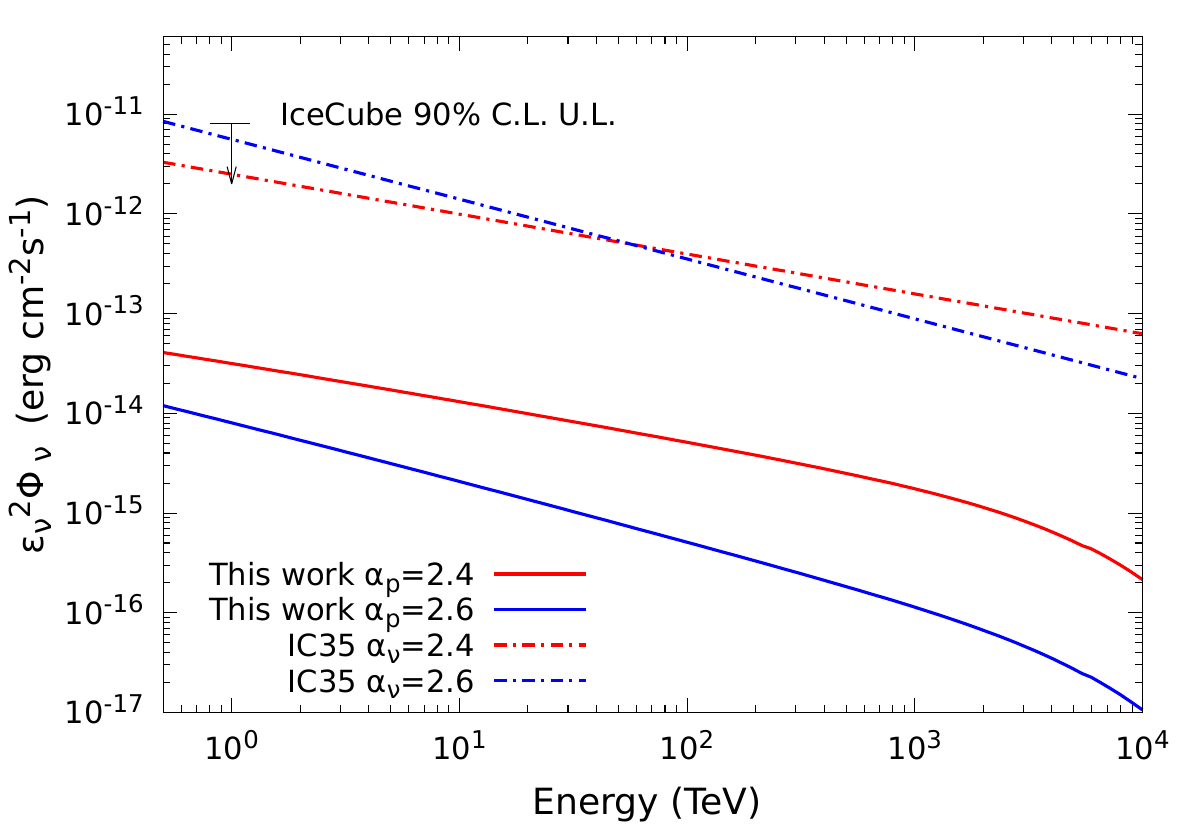}}
			}
		\caption{Neutrino flux comparison among the resulting by our model (solid-lines), the 90\% CL upper-limit for muon-neutrino at 1 TeV set by \cite{2020PhRvL.124e1103A} (black arrow),  and  the associated flux   to the IC35 event considering 10 years of observations and assuming spectral index of 2.4 (red dashed-dotted line) and 2.6 (blue dashed-dotted line). }\label{Fig_neutrino_spectrum}
\end{figure}

We calculate the expected number of neutrinos in the range of 30 TeV to 10 PeV in the IceCube detector during 10 years of observations. Using Eq. (\ref{eq_Nnu}) and an spectral index $\alpha_p$ in the range of $2.4-2.7$, we obtain the number of neutrino events to be  $N_\nu \approx (1-0.04)\times10^{-2}$. Furthermore, the neutrino flux obtained with our model is compared with that associated with the IC35 event, as shown in Fig. \ref{Fig_neutrino_spectrum}. This figure shows that fluxes are consistent with the point-source flux upper-limit established by IceCube \citep{2020PhRvL.124e1103A}, and also that our model cannot explain the IC35 flux with any value of parameters used to describe the gamma-ray observations. This implies that the IC35 neutrino event could not have been produced inside the nuclear SB region. Interestingly, the IC35 event would imply a flat spectrum instead of a steeper one as the IceCube's result for NGC 1068.


We have assumed that HN are CR accelerators up to 100 PeV. However, some authors have pointed out that these astrophysical objects could reach energies beyond EeV \citep{Wang_etal&2007PhRvD..76h3009W,Wang_etal&2008ApJ...677..432W}. Extrapolating our results at ultra-high energies,  the luminosity above 37 EeV would be $L_{\rm > 37 \, EeV} \approx 7.6 \times 10^{37} \, (7.1 \times 10^{34}) \, \rm erg \, s^{-1}$ for $\alpha = 2.4 (2.7)$.  Using these quantities in Eq. \ref{eq_L_UHECR}, we found that the expected number of UHECRs above $\rm 37 \, EeV$ would be $N_{\rm UHECR}^{\rm >37 \, EeV} \approx 1.3 (0.002)$ for $\alpha = 2.4 (2.7)$.  These values are too low to explain the anisotropy reported by Pierre Auger Observatory.

\begin{table}
\caption{Observational parameters of NGC 4945}
\begin{ruledtabular}\label{tab:ObsParameters}
\begin{tabular}{lclc}
\textbf{Starburst region} & \textbf{Value} & \textbf{Description} & \textbf{}\\
\hline
	$\rm SFR \,[M_{\odot}\, yr^{-1}] $ & 4.6  &  Star formation rate & (1) \\
	$ R \rm \, [pc]              $     & 110 & Radius & (2) \\
	
		$h \rm \, [pc]              $  & 80 & Half-scale height & (2)\\
	$R_{\rm sn} \rm \, [yr^{-1}]         $ & $>$ 0.1 & SN rate & (3)\\
	$\rm L_{FIR} [10^{43}\,erg \, s^{-1}] $ & $\rm 8$  &  Total far infrared & (2)\\ && luminosity\\

\hline
\textbf{Superwind} & \textbf{Value} & \textbf{Description} & \textbf{}\\
\hline
	$\rm \theta_{\rm sw} \,[^\circ]$ & $\rm 40$    & half-open angle & (6)\\
	$R_{\rm sw} \rm \, [kpc]        $ & $\rm \sim 1.8$   & Scale length & (4, 7)\\
	$V_{\rm sw} \rm \, [km \, s^{-1}]        $ & $\sim$ 300 - 600     & Superwind velocity & (5)\\
	$T_{\rm sw} \rm \, [keV]      $ & $\sim$ 0.6  & Temperature & (6)\\
    \end{tabular}
\end{ruledtabular}
   \\References: 
   (1) \cite{Strickland_2004},
   (2) \cite{Lipari_1997}, 
   (3) \cite{Lenc_2008}, 
   (4) \cite{2017FrASS...4...46V}, 
   (5) \cite{1990ApJS...74..833H}, 
   (6) \cite{10.1046/j.1365-8711.2002.05585.x}, and
   (7)
   \cite{1989PASJ...41.1107N}.
\end{table}
\begin{table}
\caption{Derived parameters from radio and gamma rays observations using $pp$ interaction model in the SB region}
\begin{ruledtabular}\label{tab:DerivedParameters_SB}
\begin{tabular}{lcl}
\textbf{Parameter} & \textbf{Value} & \textbf{ Description} \\
\hline
$n_{g} \rm  \, [cm^{-3}] $                & $\rm 5 \times 10^3$     & Gas number density\\
$E_p^{max} \rm  \, [PeV]       $          & $\rm \sim 100$          & Maximum proton energy\\
$\alpha      $                         & 2.4 - 2.7               & spectral index\\
$B \rm \, [ m G]       $                & 1.4 - 0.9            & Magnetic field strength\\
$L_{cr} \rm \, [10^{42} \, erg \, s^{-1}]       $ & 1.3 - 1.8$  $   & Bolometric CRs luminosity \\ & & injected by SN and HN\\
$\left< E_{\rm cr} \right> \,\rm [10^{50} \, erg]$  & 4.2 - 5.9   & Average CRs energy injected \\& &  per SN and HN\\
$R_{sn}^{\rm UL} \,\rm [yr^{-1}]$  & $\lesssim 1$   & Upper limit of SN rate\\
$N_\nu \, [10^{-2}]      $                        & $1- 0.04$       & Expected number of $\nu$ in \\ & & IceCube during 10 years \\
\end{tabular}
\end{ruledtabular}
\end{table}

\subsection{Superwind}

A conically shaped X-ray morphology was observed in NGC 4945 by \cite{10.1046/j.1365-8711.2002.05585.x}, which infers a thermal temperature of $\rm \sim 0.6 \, keV$ (similar to NGC 253). Additionally, optical emission lines were detected, attributed to a superwind with a velocity of 300-600 km/s at a distance of 70-700 pc from the nucleus \cite{1990ApJS...74..833H}. \cite{1989PASJ...41.1107N} observed optical radial filaments from a region with an extension of  $\sim $1 kpc x 6 kpc towards the halo. Along the major axis of the galaxy, these filaments are ionized gas flowing from the nuclear region.

\subsubsection{Scaling relations}

Following \cite{2009ApJ...697.2030S}, the rate of energy transferred from SNe and stellar winds to galactic wind is
\begin{equation}
\dot{E} \approx 2.5 \times 10^{41} \, \epsilon \, \rm \left( SFR / (M_\odot yr^{-1}) \right)  \; erg \, s^{-1} ,
\end{equation}
and the rate at which the hot matter is injected into the wind is
\begin{equation}
\dot{M} \approx 0.117 \, \beta \,  \, \rm \left( SFR / (M_\odot yr^{-1}) \right)  \; M_\odot \, yr^{-1}\,.
\end{equation}
From the above two quantities, it is possible to determine the temperature of hot gas in the central cavity. To be in agreement with observations performed by \cite{10.1046/j.1365-8711.2002.05585.x},  we consider the limits $0.3 \leq \epsilon \leq 1 $ obtained by \cite{2009ApJ...697.2030S} for the case of M82 and find a temperature of
\begin{equation}
T_c \approx  0.6 	\, \left( \epsilon/0.75 \right) \left( \beta/10 \right)^{-1} \rm \; keV \,.
\end{equation}
The limits of thermalization parameters provide the constrains $4 \leq \beta \leq 14 $. Using the dimensions of the starburst region, we can estimate the hot gas's number density in the central cavity as
\begin{multline}
n_c \approx 14 \, \left( \epsilon/0.75 \right)^{-1/2} \left( \beta/10 \right)^{3/2}
\times \\
\left( (h/R)/0.72 \right)^{-1} 
\left( R / 110 \, {\rm pc} \right)^{-2}\rm \; cm^{-3}. 
\end{multline}
Given the values of number densities in the range of $5 \leq n_c / {\rm cm^{-3}} \leq 20 $, the thermal pressure lies in $3 \leq P_c / ( {\rm keV \, cm^{-3}} ) \leq 12$.
%
%
%
%
Once the wind escapes from the starburst region, the terminal velocity is
\begin{equation}\label{v_term}
V_\infty  \approx 713 \, \left( {T_{c}/0.6 \rm \,  KeV} \right) ^{1/2} \rm \; km \, s^{-1}\,.
\end{equation}

\begin{table}
\caption{Derived parameters of superwind}
\begin{ruledtabular}\label{tab:DerivedParameters_SW}
\begin{tabular}{lcl}
\textbf{Parameter} & \textbf{Value} & \textbf{ Description}\\
\hline
	$\epsilon      $            & 0.3 - 1                   & Thermalization efficiency\\
	$\beta      $               & 4 -14                     & Load mass factor\\
	$n_c \, \rm [cm^{-3}]  $           & $\sim$ 5 - 20              & Density of the hot gas\\
	$V_\infty \, \rm [km \, s^{-1}]  $ & $\sim$ 713                    & Terminal velocity\\
	$L_{p} \, \rm [erg \, s^{-1}]  $   & $\sim$ $1.6 \times 10^{40} $ & Bolometric proton luminosity\\
	$E_{p}^{max} \, \rm [PeV]  $       &  $\sim$ 10                    & Maximum proton energy\\
\end{tabular}
\end{ruledtabular}
\end{table}

\subsubsection{Cosmic rays inside the Superwind}

The superwind shocks accelerate particles via the Fermi mechanism. The CR luminosity inside the superwind is $L_{\rm cr}^{\rm sw} = \xi \dot{M} V_{\rm sw}^2$ with $V_{\rm sw}$ as the terminal velocity. Taking into account Eq. (\ref{v_term}) and $\xi \approx 0.1$, the total CR luminosity becomes
\begin{multline}\label{eq_Lcr_sw}
    L_{cr}^{sw} \approx  1.6 \times 10^{40} \left( {\xi}/{0.1} \right) \left( {\rm SFR / 4.6\, M_\odot \, yr^{-1} } \right)
    \times \\
    \left( {T_c}/{0.6 \rm \, keV} \right)  \,\rm  erg \, s^{-1} \, .
\end{multline}

Considering the upper limit on the magnetic field in the halo provided by \cite{1997MNRAS.284..830E} as a value in the superwind region, the maximum proton energy that superwind can confine is
\begin{equation}
    E^{\rm max}_{\rm Z} \approx 5.6\; Z \, \left( B / \rm 6 \, \mu G  \right)^{-1} \, \left( R_{\rm sw } / {\rm  kpc} \right) \rm \; EeV.
\end{equation}
For heavy nuclei like iron, the maximum energy becomes $E_{\rm Fe}^{\rm max} \approx 1.45 \times 10^{20} \rm \, eV$. To determine if the superwind can accelerate UHECRs, we compare the dynamical and the acceleration timescales, such that $ t_{\rm acc} \lesssim t_{\rm dyn}$, where the dynamical timescale is
\begin{equation}
t_{\rm dyn}=1.37 \left({R_{\rm sw}}/{\rm \, kpc}\right) \left(V_{\rm sw}/ 700 \, {\rm km \,s^{-1}} \right)^{-1} \;\rm  Myr,
\end{equation}

and the acceleration timescale is
\begin{multline}
t_{\rm acc} \approx 5.19 \, Z^{-1} \,  \left(V_{\rm sw}/ 700 \, {\rm km \,s^{-1} }\right)^{-2} 
\left( D/D_{\rm B}\right) 
\times \\
\left( {E}_{\rm p} / 100 \, {\rm PeV} \right) 
\left( B / \rm 8 \, \mu G  \right)^{-1} \; \rm Myr \,.
\end{multline}

Taking into account the relation $t_{\rm dyn} \approx t_{\rm acc}$, the maximum proton energy reached is $E^{\rm max}_{\rm p} \approx 10 \; \rm PeV$, and in the case of iron nuclei is $E^{\rm max}_{\rm Fe} \approx 350 \; \rm PeV$. Therefore, superwind in NGC 4945 cannot explain the Auger hotspot above $\rm 39 \, EeV$.
Our result agrees with that found for the similar SBG NGC 253 \cite{2018A&A...616A..57R}. On the other hand,  our result is different from the conclusion reported by \cite{Anchordoqui_2020} because we use a more conservative assumption while they used the larger size and larger magnetic field in the Superwind.  Moreover, if the Superwind could accelerate UHECRs, as pointed out by \cite{Anchordoqui_2020}, then the luminosity above 37 EeV would be $L_{>\rm 37 \, EeV} \approx 1.5 \times 10^{40}$,   $1.2 \times 10^{38}$ and  $9.3 \times 10^{35} \, \rm erg \, s^{-1}$ for spectral indexes of $\alpha = 2$, $2.2$ and $2.4$, respectively. These luminosities correspond to $N_{\rm UHECR}^{\rm > 37 \, EeV} \approx 7.6$, $1.1$ and $1.4\times 10^{-2}$ UHECRs, which are too low explain the 62 events over the expected background \cite{2019ICRC...36..206C}.  From Eq. \ref{eq_Lcr_sw}, we can note that by increasing the star formation rate, the CR luminosity could be enhanced by a factor of $\sim 3$ considering the gamma-ray observations.  Therefore, the number of UHECRs could be increased by the same factor reaching a significant fraction of the anisotropic flux, i.e.,  $N_{\rm UHECR}^{\rm > 37 \, EeV} \sim 22$, but only for the case of $\alpha = 2$. 
Finally, protons at 10 PeV interact with the hot gas inside the superwind with a timescale of
\begin{equation}
 t_{pp} \approx 30 \, \left( {n_{g}}/{\rm 10^{-2} \, cm^{-3}} \right)^{-1} \left( {\sigma_{ pp}}/{\rm 70 \, mb} \right)^{-1} \; \rm Gyr\,.
\end{equation}
Therefore, the efficiency of $pp$ collision in the halo region is very low $f_{\rm pp}\sim 10^{-4}$, implying not a very significant contribution to neutrinos and gamma-rays in comparison with the central SB region. Our result of the Superwind region is summarized in Table \ref{tab:DerivedParameters_SW}.

\subsection{Core region}
 
The association of the neutrino hotspot with the Seyfert 2 galaxy NGC 1068 has been discussed in the scenario of the AGN coronae-disk model by \cite{2021arXiv210204475K,2020ApJ...891L..33I}.  In this model, neutrinos and MeV gamma-rays emerge, and gamma-rays above GeV are fully attenuated by photons from the UV-disk and X-ray corona. The issue with this scenario is that PeV neutrinos are difficult to explain because protons cannot accelerate beyond PeV energies due to cooling processes. In the case of NGC 4945, no significant neutrino emission at TeV energies was found by \citep{2020PhRvL.124e1103A}, and the spectrum associated is harder than the one seen for the NGC 1068 case.
 
NGC 4945 has an accretion disk with a temperature close to $T_d = 10^4 \, \rm K$, which radiates UV-photons at an energy of $\epsilon_{d} \approx 2.3 \, \rm eV$.   Protons with energies above $\epsilon_{p,\rm th}^{p\pi}  \gtrsim 30  \, {\rm PeV} \, (\epsilon_d/2.3\, {\rm eV})^{-1}$ interact with UV-photons via photopion processes and produce PeV neutrinos above $\epsilon_\nu \gtrsim 1.6 \, {\rm PeV} \, (\epsilon_d/2.3\, {\rm eV})^{-1}$. The corresponding photopion efficiency could be estimated as $f_{p\pi} \approx \sigma_{p\pi} \, R \, n_d \approx 1 \, (R/{10^{14} \, \rm  cm}) (T_d/ 10^4 \, \rm K)^3$, where the size is restricted by the variability timescale of the order of $\sim 10^4 \, \rm sec$. Therefore, the key to produce PeV neutrinos in the AGN-core is to accelerate protons beyond $\sim \rm PeV$ energies near the accretion disk. These protons could be accelerated by a relativistic jet, as observed in NGC 1068. Similarly, NGC 4945 exhibited a structure with a morphology type, suggesting a core-jet with a size of 5 pc of length and 1.5 pc of width \cite{Lenc_2008}.

\section{Conclusion}
SBGs are promising sources of HE neutrinos as well as UHECRs. We analyzed if the nearby SBG NGC 4945 could contribute to the anisotropic region reported by Pierre Auger Observatory, and be associated with the neutrino event IC35, which is close to this region. We considered two regions around this galaxy: the SB and the superwind or halo region.  

For the SB region, we used spectral indexes in the range of 2.4 and 2.7 to explain the Fermi and radio observations. We found that the average CRs energy injected per SN or HN is $\left< E_{\rm CR} \right> \approx 4.2-5.9 \times 10^{50} \rm \, erg$. Our estimations show that HN and SN have similar contributions in the luminosity of CRs and similar CR efficiencies $\eta_{sn,hn} \approx 0.3$. Also, if there is not CR contribution by HNe, an upper value of SN rate can be set $R_{sn} \lesssim 1 \, \rm yr^{-1}$.  This estimate is lower than the value reported by other authors.   Additionally, we found that SB provides at a proton luminosity at least $\sim 100$ times larger than that generated in the Superwind. The total gamma-ray flux is estimated by ${\rm pp}$ interactions in the SB region,  by bremsstrahlung, and from secondary electrons, i.e., pionic and pairs produced in the interactions of primary gamma-rays and the SB's radiation field.  The radio observation is explained by both the primary and secondary pionic electrons, but demanding a magnetic field of $\sim 1 \, \rm mG$.

We estimated the maximum energy that CRs could reach in the nuclear SB and the superwind regions.  In the nuclear SB region, particles are accelerated by HN reaching energies of $\rm \sim 100 \, Z \, PeV$.  We extrapolated the CR spectrum obtained by modelling gamma-rays up to ultra high energies and showed that the contribution to the hotspot is too low (less than one event). 

For the Superwind case, protons are accelerated in the shock front via Fermi acceleration.  The maximum energy that could be reached is $\rm \sim 10 \, Z \, PeV$, although this value could be higher if we do not consider conservative parameter values. If the Superwind accelerates UHECRs, the contribution to the Auger's hotspot is $\sim 7$ events if we assume a flat spectrum ($\alpha=2$). This value could be enhanced near one-third of the total number of events if we increase the value of SFR without overproducing the gamma-rays spectrum. This result suggests that the SBG  NGC 4945 could not be the only one responsible for the Auger hotspot.   Therefore, we concluded that only a tiny fraction of UHECRs reported by Pierre Auger Observatory could have been produced in NGC 4945 (by either the SB region or the superwind). It leads to the open question about the origin of UHECRs.

Finally, we estimated the expected number of neutrinos in the energy range from 30 TeV to 10 PeV from the SB region. We found that this number lies in the range of $(0.04-0.1)\times10^{-2}$ for ten years of IceCube observations. Also, we showed that IC35 flux was consistent with the upper limit set by IceCube. Therefore, we concluded that the SB region could not produce the IC35 event. It is worth noting that our conclusion does not discard the  AGN-core, which, as discussed above, is a highly efficient region for $\gtrsim$ PeV-neutrino production. However, a feasible mechanism to accelerate CRs above $\gtrsim$ 10 PeV is unclear for Seyfert galaxies, e.g., NGC 1068 case.  If the AGN-core of NGC 4945 produced the IC35 event, then a flat neutrino spectrum would be associated to it. If this is the case,  significant differences must exist with the neutrino emission mechanism in NGC 1068.  An opposite case could be happening in  NGC 1068, where a possible association with a TeV-neutrino hotspot could exist. In this case, the absence of a PeV neutrino would indicate a steeper spectrum, i.e., $\phi_{\nu_\mu}\propto E^{-3.2}$.  Therefore, detailed studies are needed in the AGN-core of these two SB/Seyfert galaxies to claim differences in PeV-neutrino emission and CR acceleration signatures.

\section{Acknowledgements}
We thank Kotha Murase, Luis Anchordoqui, Ana Laura Müller, Floyd Stecker and Gustavo Romero for useful discussions.  This work is supported  by UNAM-DGAPA-PAPIIT  through  grant  IN106521.

%
    
\begin{figure}
		{
        \resizebox*{0.5\textwidth}{0.28\textheight}
		{\includegraphics{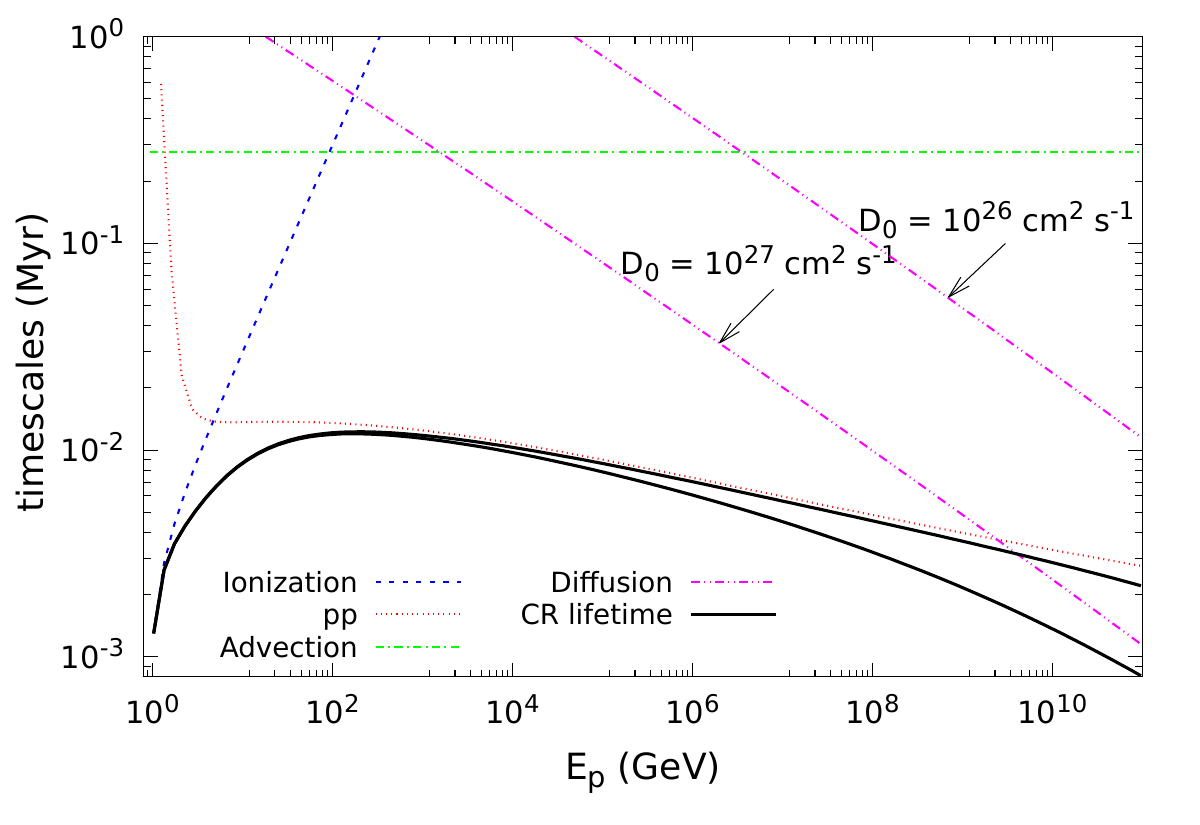}}
		}
		\caption{CR proton timescales: Losses via ionization and inelastic $pp$ interaction are taken. The escape is due to advection and diffusion mechanism. }\label{fig_p_timescales}
\end{figure}

\appendix
\section{CR timescales}
\subsection{CR escape timescale}
We have discussed below the numerical values of characteristic timescales, which affect the CR distribution in the SB region. The escape time is a competition between diffusion and advection $t_{\rm  esc}^{-1} = t_{\rm  diff}^{-1} + t_{\rm  adv}^{-1}$. Electrons and protons escape following the same reaction.  Assuming a Kolmogorov turbulence spectrum as observed in our Galaxy,  the diffusion coefficient is parametrized as 
$D(E_p) = D_0 \, (E_p/3 \, \rm GeV)^{0.3}$ where $D_0$ is the diffusion coefficient  at 3 GeV. The value of $D_0$ in NGC 4945 must be lower than the one obtained by our Galaxy, i.e., $D_0 \sim 6 \times 10^{28} \, \rm cm^2 \, s^{-1}$ \citep{2011ApJ...729..106T} . This assumption is because of SBs galaxies have larger magnetic field strength than our galaxy, and the diffusion coefficient is expected to scale with the Larmor radius as $D\propto r_L$. Here, we use the assumption $D_0 \lesssim 10^{27} \, \rm cm^2 \, s^{-1}$ as done in \citep{2014PhRvD..89h3004L}. Therefore, the diffusion time is given by
\begin{multline}
t_{\rm diff} 
=  \frac{ 3 h^2}{4D} 
\\ \approx 
10 \, \left( \frac{h}{\rm 80 \, pc} \right)^2 \left( \frac{E_p}{\rm 50 \, PeV} \right)^{-0.3} \left( \frac{D_0}{\rm 10^{27} \, cm^2 \, s^{-1} } \right)^{-1} \; \rm kyr \, .
\end{multline}

Another way that proton can escape is via advective transportation by the galactic wind. Considering the superwind lowest velocity the timescale is given \begin{equation}
t_{\rm adv} = \frac{h}{V_w} \approx 260 \, \left( \frac{h}{\rm 80 \, pc} \right) \left( \frac{V_w}{\rm 300  \, km/s} \right)^{-1} \; \rm kyr \, .    
\end{equation}
 From above estimation, we notice that advection is the principal electron's escape way and the diffusion mechanism can be neglected.
\subsection{Proton loss timescales}
The proton lifetime is given by 
\begin{equation}
\tau_{p} = \left( t_{\rm ion,p}^{-1} + t_{\rm  pp}^{-1} + t_{\rm esc}^{-1}\right)^{-1}\,.    
\end{equation}%
At low energies,  proton mainly losses energy by ionization \cite{Schlickeiser2002}
\begin{multline}
    t_{{\rm ion},p}^{-1}= 1.82 \times 10^{-7} \; E_p \, n_g \,
    \frac{2\beta^2}{10^{-6}+2\beta^3} 
    \\
    \left[ 1 + 0.0185\ln(\beta) \Theta(\beta-0.01) \right]
    \\
    \approx 
    1 \left( \frac{E_p}{\rm \, GeV} \right) \left( \frac{n_g}{\rm \, 5 \times 10^3 \, cm^{-3}} \right) \; \rm kyr^{-1}, \;  \quad 
\end{multline}
where $\beta$ is the proton velocity and $\Theta$ is the Heaviside function.
The energy loss timescale via $pp$ collision is simple approximated as
\begin{multline}
t_{pp}^{-1}  \simeq \kappa	 \, \sigma_{pp}(E_p) \, c \, n_{g} \Theta(E_p - 1.22 \rm GeV)
\\
\approx 0.81
1 \left( \frac{\sigma_{pp}({\rm TeV})}{\rm 34 \,mb} \right) \left( \frac{n_g}{\rm \, 5 \times 10^3 \, cm^{-3}} \right) \, \rm kyr^{-1} \, .
\end{multline}

%
%
From comparing the above timescales, we can infer that the advection process dominates protons' escape while losses dominate the total lifetime via $pp$ interactions. Furthermore, the re-acceleration processes can be neglected. The proton timescales are calculated as energy's function and are plotted in Figure \ref{fig_p_timescales}.

\subsection{Electron loss timescale}
Electrons are accelerated and injected together protons, and the principal energy losses are synchrotron, IC, Bremsstrahlung, and ionization. Then, electron lifetime is given by  
\begin{equation}
\tau_{e} = \left( t_{\rm ion}^{-1} + t_{\rm  brem}^{-1} + t_{\rm syn}^{-1} + t_{\rm IC}^{-1} + t_{\rm esc}^{-1}\right)^{-1}    
\end{equation}
Considering an electron Lorentz factor of $\gamma_e \approx 10^3$ as typical break value, we calculate the losses timescale (e.g,  \cite{2013MNRAS.430.3171L} and references therein). The synchrotron cooling time is
\begin{multline}
t_{\rm syn}^{-1} = \frac{\sigma_T \, c \, B^2 \gamma_e}{6\pi m_e c^2}
\\
\approx 0.092 \left( \frac{\gamma_e}{10^3} \right) \left( \frac{B}{\rm 750 \, \mu G} \right)^2 \; \rm kyr^{-1},    \qquad\quad
\end{multline}

the IC cooling time in the Thompson regime \citep[$\gamma_e \epsilon \lesssim m_e c^2$;][]{2014MNRAS.441.1209F,2017APh....89...14F} is given by
\begin{multline}
t_{\rm IC}^{-1} = \frac{4 \sigma_T \, c \, U_{ph} \gamma_e}{3m_e c^2}
\\
\approx  0.036\left( \frac{\gamma_e}{10^3} \right) \left( \frac{U_{\rm IR}}{\rm 10^{-8} \, erg \, cm^{-3}} \right) \; \rm kyr^{-1},  
\end{multline}

the Bremmstrahlung timescale is only dependent on the gas density \citep{Longair2011}
\begin{multline}
t_{\rm brems}^{-1} = n_g \frac{Z^2e^6}{12\pi^3m_e^2 \epsilon_0 c^4 h} \ln \left(192/Z^{1/3}\right)
\\
\approx  0.128 \left( \frac{n_g}{\rm 5 \times 10^3 \, cm^{-3}} \right) \; \rm kyr^{-1}, \qquad\quad \;
\end{multline}

while the ionization process is energy dependent
\begin{multline}
t_{\rm ion}^{-1} = {\gamma_e}^{-1} \left[ 2.7 \, c \, \sigma_T \left( 6.85 + 0.5 \ln\gamma_e \right) n_g   \right]
\\
\approx 0.023 \left( \frac{\gamma_e}{10^3}\right)^{-1} \left( \frac{n_g}{\rm 5 \times 10^3 \, cm^{-3}}  \right) \; \rm kyr^{-1}  \,.
\end{multline}

All loss processes are comparable at GeV energies, and none can be neglected.

\bibliography{apssamp}

\end{document}